\documentclass[12pt,preprint]{aastex}
\usepackage{psfig,epsfig,amsfonts,amsmath,amssymb,graphicx,morefloats,appendix,tensor}
\usepackage{color}
\usepackage{natbib}
\begin{document}

\slugcomment{Accepted by ApJ: July 6, 2015}

\title{
The Epsilon Eridani System Resolved by Millimeter Interferometry
}

\author{Meredith A. MacGregor, David J. Wilner, Sean M. Andrews}
\affil{Harvard-Smithsonian Center for Astrophysics}
\affil{60 Garden Street, Cambridge, MA 02138, USA}
\email{mmacgregor@cfa.harvard.edu}
\and
\author{Jean-Fran\c{c}ois Lestrade}
\affil{Observatoire de Paris - LERMA, CNRS}
\affil{61 Av. de l'Observatoire, F-75014, Paris, France}
\and
\author{Sarah Maddison}
\affil{Centre for Astrophysics \& Supercomputing}
\affil{Swinburne University, Melbourne, Australia}

\begin{abstract}
We present observations of $\epsilon$ Eridani from the Submillimeter Array 
(SMA) at 1.3~millimeters and from the Australia Telescope Compact Array 
(ATCA) at 7~millimeters that reach an angular resolution of $\sim 4\arcsec$ (13~AU). 
These first millimeter interferometer observations of $\epsilon$~Eridani,
which hosts the closest debris disk to the Sun, reveal two distinct emission 
components: (1) the well-known outer dust belt, which, although patchy, is clearly resolved 
in the radial direction, and (2) an unresolved source coincident with the 
position of the star. We use direct model-fitting of the millimeter 
visibilities to constrain the basic properties of these two components. 
A simple Gaussian shape for the outer belt fit to the SMA data results in a 
radial location of $64.4^{+2.4}_{-3.0}$~AU and FWHM of $20.2^{+6.0}_{-8.2}$~AU 
(fractional width $\Delta R/R =0.3$). Similar results are obtained taking a 
power law radial emission profile for the belt, though the power law index 
cannot be usefully constrained. Within the noise obtained 
(0.2~mJy~beam$^{-1}$), these data are consistent with an axisymmetric belt 
model and show no significant azimuthal structure that might be introduced by 
unseen planets in the system.  
These data also limit any stellocentric offset of the belt to $<9$~AU, 
which disfavors the presence of giant planets on highly eccentric ($>0.1$) 
and wide (10's of AU) orbits.
The flux density of the unresolved central component exceeds predictions 
for the stellar photosphere at these long wavelengths, by a marginally
significant amount at 1.3~millimeters but by a factor of a few at 
7~millimeters (with brightness temperature $13000\pm1600$~K for a source
size of the optical stellar radius). We attribute this excess emission to 
ionized plasma from a stellar corona or chromosphere.
\end{abstract}


\section{Introduction}

Debris disks, composed of planetesimals remaining after planet formation 
and circumstellar disk dispersion, represent the end-stage of protoplanetary
disk evolution \citep[see reviews by][]{bac93,wya08,mat14}.  While these 
remnant planetesimals cannot be observed directly, they are ground down 
through ongoing collisions into smaller and smaller dust grains that scatter 
starlight and produce detectable thermal emission. Observations of this dusty 
debris at millimeter wavelengths are especially critical to our understanding 
of the most readily accessible systems.  The large grains that dominate 
emission at these long wavelengths do not travel far from their origin and 
therefore reliably trace the underlying planetesimals distribution, unlike 
the small grains that are rapidly removed by stellar radiation and 
winds \citep{wya06}.  Since planets, if present, will inevitably perturb 
the dust-producing planetesimals, the millimeter emission morphology encodes 
information on the architecture and dynamical evolution of these systems. 
For example, the outward migration of a planet can confine planetesimals in
a belt between its resonances \citep{hah05}, or trap planetesimals into 
mean motion resonances outside its orbit \citep{kuc03,wya03,del05}. 
A planet can also sculpt out sharp edges in a belt \citep{qui06,chi09}, or 
force planetesimals onto eccentric or inclined orbits \citep{wya99}.  

At a distance of only 3.22~pc \citep{vanL07}, 
the $400-800$ Myr-old \citep{mam08}
main-sequence K2 star $\epsilon$ Eridani hosts the closest debris disk to the
Sun, originally identified through the detection of far-infrared emission by
{\em IRAS} \citep{aum85}.  Pioneering observations
with JCMT/SCUBA resolved a nearly face-on belt of emission at 850~$\mu$m,
peaking at 60 AU ($18\arcsec$) radius, with several brightness enhancements
or clumps \citep{gre98}.  Analysis by \cite{gre05} of JCMT images spanning
5 years offered tentative evidence that some of these clumps are stationary,
and so likely background galaxies, while others appear to be co-moving with
the star, and so are likely associated with the disk.  Subsequent single-dish observations 
from 250 to 1200 $\mu$m have confirmed the basic belt morphology, but not all 
of the low significance asymmetries \citep{sch04,bac09,gre14,les15}.
A warmer dust component, reaching to several AU from the star, can be
explained in modeling the spectral energy distribution by an additional dust belt 
\citep{bac09,gre14}, or by inward transport from the outer belt \citep{rei11}.
In addition, precision radial velocity observations suggest the presence
of a Jupiter-mass planet with semi-major axis of 3.4 AU ($1''$) \citep{hat00},
although the reality of this planet signal remains controversial \citep{ang12}.
Because $\epsilon$~Eridani is so nearby, it is a key template for
understanding debris disk phenomena around Sun-like stars, and detailed
study of its debris disk provides essential context for the interpretation
of more distant, less accessible systems.

We present observations of $\epsilon$ Eridani at 1.3~mm and at 7~mm, using 
the Submillimeter Array (SMA) and the Australia Telescope Compact Array (ATCA), 
respectively, using the most compact, lowest angular resolution ($4''-10''$) 
configurations of these telescopes. While even higher resolution may be 
desirable, these interferometric observations are conservatively tuned to 
provide a first
look at structures below the resolution of previous single dish observations. 
For these arrays at these wavelengths, the primary beam field of view 
encompasses the entire emission region from the outer debris belt, enabling 
efficient observations of the full disk with a single pointing.
Section~\ref{sec:obs} describes these observations of the $\epsilon$ Eridani 
system. Section~\ref{sec:analysis} describes the modeling procedure and the 
results. Section~\ref{sec:discussion} discusses the implications of the 
model fits for the outer dust belt properties, azimuthal asymmetries, and 
the nature of an inner component of excess millimeter emission.

\section{Observations}
\label{sec:obs}

\subsection{Submillimeter Array}

We observed $\epsilon$ Eridani in July, August and November 2014 
with the SMA \citep{ho04} on Mauna Kea, Hawaii at a wavelength of 1.3~mm 
in the subcompact configuration. Table~\ref{tab:sma} summarizes the 
essentials of these observations, including the dates, baseline lengths, 
and atmospheric opacity.  Six tracks were obtained, all with 7 operational 
antennas in the array.  The weather conditions were very good for 
observations at this wavelength (225~GHz opacities from 0.07 to 0.12).
The total bandwidth available was 8 GHz consisting of two sidebands of 4 GHz width 
spanning $\pm4$ to 8 GHz from the local oscillator (LO) frequency of 
225.5 GHz ($217.5 - 221.5$ GHz and $229.5 - 233.5$ GHz). The phase center was 
located at $\alpha = 03^\text{h}32^\text{m}54\fs9024$, 
$\delta = -09\degr27\arcmin29\farcs4486$ (J2000), corresponding to 
the position of the star corrected for its proper motion of 
(-975.17,19.49) mas~yr$^{-1}$ \citep{vanL07}  as of July 1, 2014 .  
At the LO frequency, the field of view is $\sim 52 \arcsec$, set by the primary beam size of the 6-m diameter array antennas.  

The data from each track were calibrated independently using the IDL-based 
MIR software package.  Time-dependent complex gains were determined from
observations of two nearby quasars, J0339-017 ($7\fdg9$ away) and J0423-013 
($14\fdg9$ away), interleaved with observations of $\epsilon$ Eridani in 
a 16 minute cycle. The passband shape was calibrated using available bright 
sources, mainly 3C84 or 3C454.3.  Observations of Uranus or Callisto 
during each track were used to derive the absolute flux scale with an 
estimated accuracy of $\sim10\%$.  
Imaging and deconvolution were performed with the {\tt clean} task 
in the CASA software package. A variety of visibility weighting schemes 
were used to explore compromises in imaging between higher angular 
resolution and better surface brightness sensitivity.  With natural weighting, 
the beam size is $6\farcs0 \times 5\farcs5$ ($19 \times 18$ AU) and the rms 
noise level is 0.17~mJy~beam$^{-1}$. The longest baselines in the dataset 
probe size scales of $\sim4''$ (13 AU).

\subsection{Australia Telescope Compact Array}

We observed $\epsilon$ Eridani in late June and early August 2014 with the 
ATCA, located near 
Narrabri, NSW, at a wavelength of 7~mm using the compact H75 and H168 
configurations of the array.  Table~\ref{tab:atca} 
summarizes the essentials of these observations.  Four tracks were obtained in 
each of the two antenna configurations, all with 6 operational antennas.  
The winter weather provided good atmospheric phase stability for this ATCA
high frequency band (rms path typically $<150$ microns from the seeing 
monitor), especially for the short baselines of interest, except for the last 
track in the H168 configuration (2014 July 28), which proved to be unusable 
due to high winds and relatively poor seeing.  Data from the 
stationary sixth antenna of the array, located $\sim 6$~km from the others, 
was discarded, given the large gap from the rest of the antennas and less
stable phase on the much longer baselines.
The bandwidth provided by the Compact Array Broadband Backend was 8 GHz, 
with 2 GHz wide bands centered at 43 GHz and at 45 GHz, in two polarizations \citep{wils11}.  
The phase center was identical to the contemporaneous SMA observations.
The field of view is $\sim 70 \arcsec$, set by the primary beam size of 
the 22-m diameter array antennas. 

The data from the seven usable tracks were calibrated independently using 
the miriad software package.  Time-dependent complex gains were determined 
using the nearby quasar 0336-019 ($7\fdg8$ away), interleaved with 
observations of $\epsilon$ Eridani in a 12 minute cycle. The passband shape 
was calibrated using the available bright sources 1921-293 or 0537-441.  
Observations of 1934-638 and Uranus during each track were used to derive 
the absolute flux scale. Comparison of the derived fluxes for 0336-019 
for each of the seven nights shows a maximum difference of 6\%, and we 
conservatively estimate the flux scale accuracy is better than $10\%$.
Imaging and deconvolution were performed with the standard routines 
{\tt invert}, {\tt clean}, and {\tt restor} in the miriad software package.  

\section{Results and Analysis}
\label{sec:analysis}

\subsection{Continuum Emission}

\begin{figure}[ht]
%\vspace{-1.0truecm}
% \hfill
\begin{minipage}[h]{0.36\textwidth}
  \begin{center}
       \includegraphics[scale=0.5]{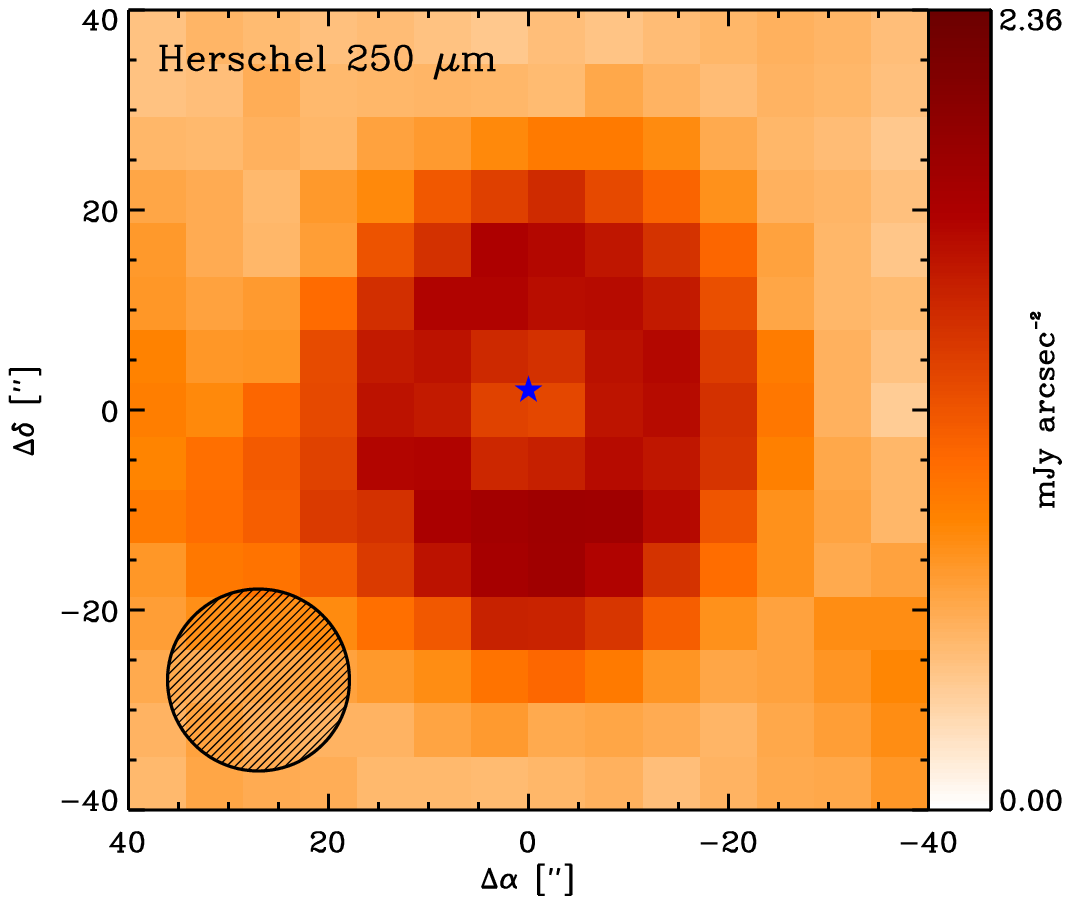}
  \end{center}
 \end{minipage}
 \begin{minipage}[h]{0.3\textwidth}
  \begin{center}
       \includegraphics[scale=0.5]{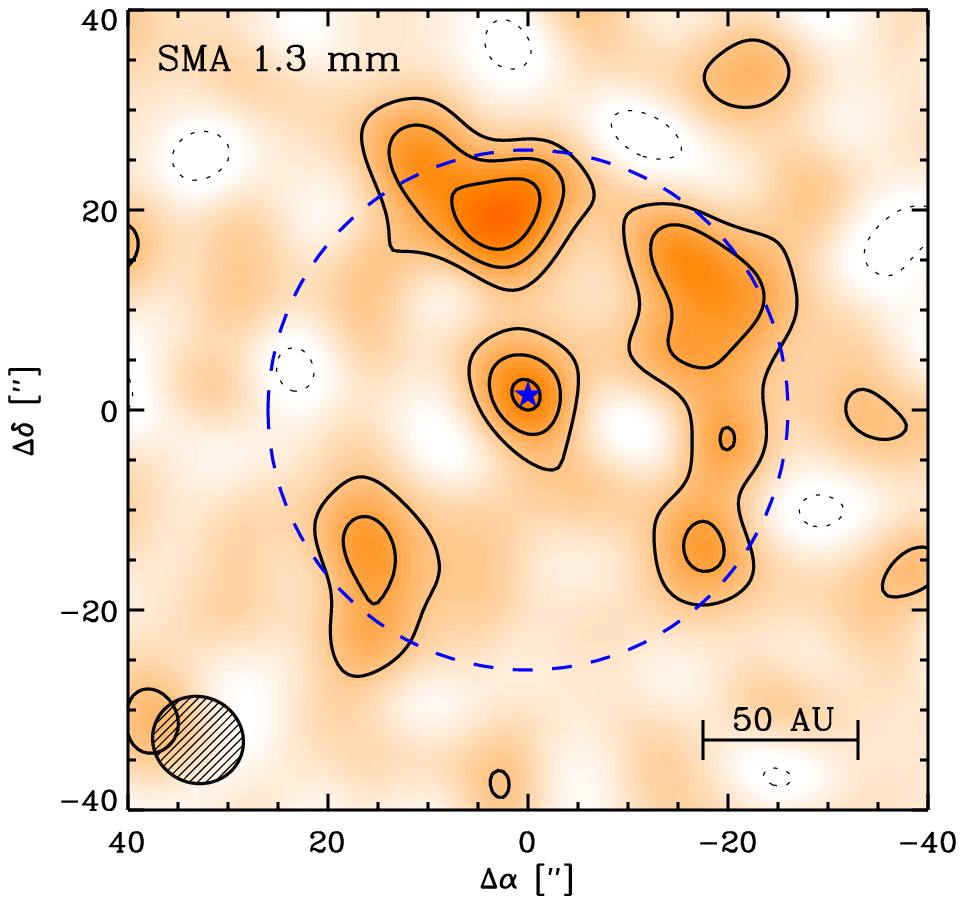}
  \end{center}
 \end{minipage}
% \hfill
  \begin{minipage}[h]{0.3\textwidth}
  \begin{center}
       \includegraphics[scale=0.5]{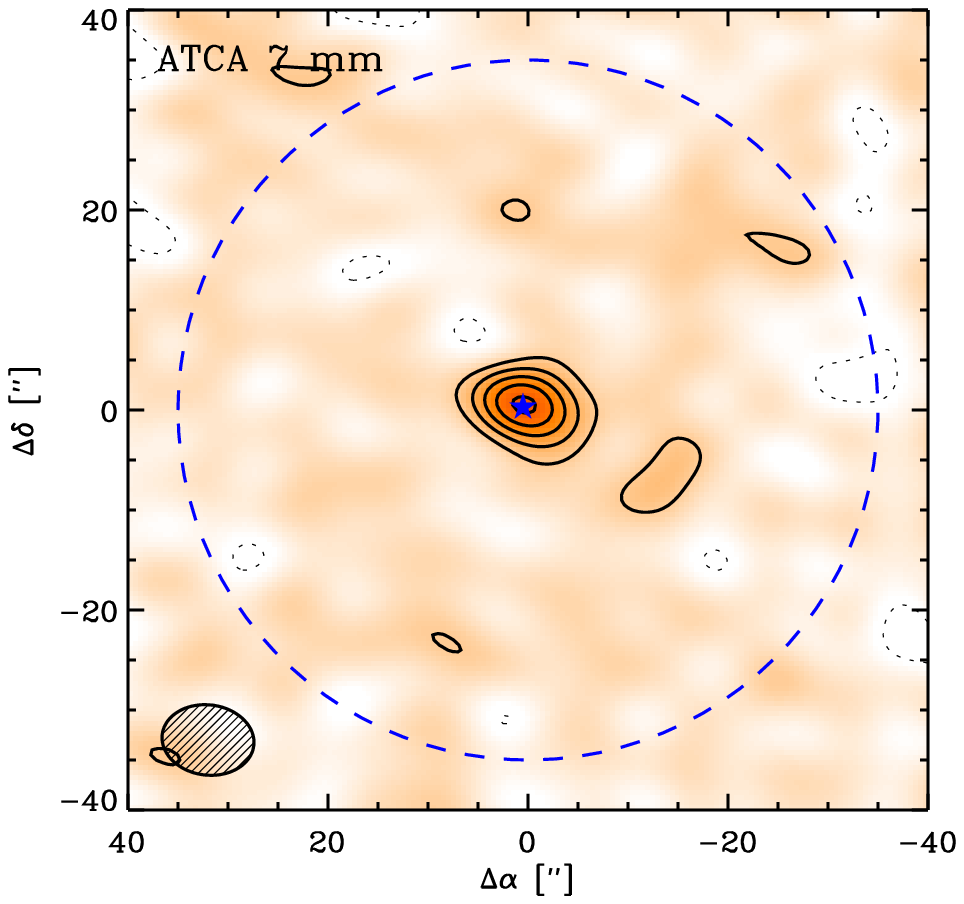}
    \end{center}
  \end{minipage}
%  \hfill
\caption{\small \emph{(left)} \emph{Herschel/SPIRE} 250~$\mu$m image of 
$\epsilon$ Eridani, from the Herschel Science Archive 
\citep[see also][]{gre14}.  
The ellipse in the lower left indicates the $\sim 19\arcsec$ beam size. \emph{(center)} SMA image of the 1.3 millimeter continuum 
emission from $\epsilon$~Eridani.  The contour levels are in steps of 
$2 \times 0.2$ mJy, the rms noise level.  The ellipse in the lower 
left corner indicates the $9\farcs2 \times 8\farcs7$ (FWHM) synthesized beam 
size. The dashed blue circle indicates the 
$\sim52\arcsec$ SMA primary beam (FWHM) at the LO frequency (225.5 GHz).
\emph{(right)} ATCA image of the 7 millimeter continuum emission from 
$\epsilon$~Eridani.  The contour levels are in 
steps of $2 \times 7$ $\mu$Jy, the rms noise level.  The ellipse in the lower 
left corner indicates the $9\farcs2 \times 7\farcs0$ (FWHM) synthesized beam 
size. The dashed blue circle indicates the 
$\sim70\arcsec$ ATCA primary beam (FWHM) at 44 GHz.
In all three panels, the star symbol marks the position of the stellar photosphere, corrected
for proper motion. For the center and right panels, the stellar position is: $\alpha = 03^\text{h}32^\text{m}54\fs9024$, 
$\delta = -09\degr27\arcmin29\farcs4486$ (J2000).
}
\label{fig:SMA_ATCA}
\end{figure}

Figure~\ref{fig:SMA_ATCA} shows an SMA 1.3~mm image and ATCA 7~mm image of
$\epsilon$~Eridani, together with the {\em Herschel/SPIRE} 250~$\mu$m image 
extracted from the Herschel Science Archive for reference \citep{gre14}. 
For the 1.3~mm image, the synthesized beam size, obtained with natural 
weighting and a modest taper to improve surface brightness sensitivity is 
$9\farcs2 \times 8\farcs7$ ($30 \times 28$ AU), position angle $68\degr$.  
The rms noise is 0.20 mJy~beam$^{-1}$. This image reveals emission from a
compact central source at the stellar position ($\sim 7\sigma$) together 
with patchy emission from the nearly face-on belt of cold dust located 
$\sim18''$ from the star.
For the 7~mm image, the synthesized beam size obtained with natural weighting 
is $9\farcs2 \times 7\farcs0$ ($30 \times 23$ AU), position angle $83\degr$.  
The rms noise is $7$~$\mu$Jy~beam$^{-1}$. A central peak is clearly detected 
($\sim 10\sigma$). Unlike the 1.3~mm image, the 7~mm image shows little sign 
of emission from the outer dust belt.
In both images, the position of the central peak is consistent with 
the predicted stellar position, within the uncertainty dictated by the 
synthesized beam size, $\theta$, and the signal-to-noise ratio, $SNR$ 
of $\sim0.5 \theta /{SNR} \approx 0\farcs6$ \citep[see][]{rei88}.

\subsection{Emission Modeling Procedure}

To characterize the 1.3 mm emission from $\epsilon$ Eridani, we used the 
procedure described by \cite{mac13,mac15} that employs a 
Markov Chain Monte Carlo (MCMC) method to fit simple parametric models to 
the observed visibilities.  We fit the visibility data directly both 
to avoid the non-linear effects of deconvolution and to take advantage 
of the full range of spatial frequencies in the observations that are not
necessarily represented well in the images.
We assume that the emission arises from a geometrically thin, 
axisymmetric belt, where we consider two different parametric shapes for the 
surface brightness profile, $I_\nu(r)$: 
(1) an annulus with $R_\text{in} < r < R_\text{out}$ and power law slope, 
$I_\nu(r) \propto r^x$, and 
(2) a Gaussian, 
$I_\nu(r) \propto \text{exp}[-((r-R_\text{cen})/\sqrt{2}\sigma)^2]$, 
where $R_\text{cen}$ is the position of the belt, $\sigma$ is the width, and 
the FWHM $= \Delta R =  2\sqrt{2\text{ln}(2)}\times\sigma$. 
The limited signal-to-noise of the dataset precludes exploring more 
complicated, but physically plausible, surface brightness shapes, such as 
multiple rings or a broken power law. 
The belt emission normalization is defined by a total flux density,
$F_\text{belt} = \int I_\nu d\Omega$, 
and the belt center is given by offsets relative to the pointing center 
$\{\Delta\alpha,\Delta\delta\}$.  
The central peak coincident with the stellar position is described by a 
point source with total flux, $F_\text{cen}$, and the offsets of this 
point source relative to the belt center are given by two additional 
parameters $\{\Delta\alpha_\text{star},\Delta\delta_\text{star}\}$.
The previous imaging observations show that the belt is viewed close 
to face-on ($i = 30\degr$, \citealp{gre14}). We fit the SMA data directly 
for an inclination angle, $i$, and also an orientation on the sky described 
by a position angle, $PA$, east of north.

The $\epsilon$ Eridani disk spans a large angle on the sky, approaching
(or possibly exceeding) the half power field of view of the SMA. 
As a result, the primary beam 
response has the potential to affect the properties derived for the outer
regions of the millimeter emission belt. To account for this in the analysis,
we multiply each belt model by an accurate, frequency-dependent beam model, 
normalized to unity at the beam center.  Appendix A provides a detailed 
discussion of the SMA primary beam shape.  For ATCA, with its larger field 
of view at the observed wavelength, the effects of the primary beam shape 
are much less important, and a simple Gaussian provides an adequate 
description.

For each set of model parameters, we use the miriad {\tt uvmodel} task 
to compute two synthetic visibility sets sampled at the same spatial 
frequencies as our SMA observations, corresponding to the two spectrally 
averaged sidebands (218.9 and 230.9 GHz). The fit quality is characterized by a 
likelihood metric, 
$\mathcal{L}$, determined from the $\chi^2$ values computed using the real 
and imaginary components at all spatial frequencies 
(ln$\mathcal{L} = -\chi^2/2$).  This modeling scheme is implemented using the 
affine-invariant ensemble MCMC sampler proposed by \cite{goo10} and realized 
in \texttt{Python} by \cite{for13}.  A MCMC approach allows us to 
more effectively characterize the multidimensional parameter space of this 
model and to determine the posterior probability distribution functions for 
each parameter.  We assumed uniform priors for all parameters, with reasonable 
bounds imposed to ensure that the model was well-defined: 
$F_\text{belt} \geq 0$ and $0\leq R_\text{in} < R_\text{out}$.  
In addition, we constrained the four offset parameters 
$\{\Delta\alpha,\Delta\delta,
\Delta\alpha_\text{star},\Delta\delta_\text{star}\}$ that 
describe the belt center and stellar position to be within $5\arcsec$ of 
the offsets predicted from the stellar proper motion; this constraint very 
generously accommodates the uncertainties in the proper motion and the 
absolute astrometry of the observations. 

For the ATCA 7~mm observations, where the $\epsilon$ Eridani emission belt is 
not visible in the image, we simplified the model substantially. We fix all 
of the model parameters to the best-fit values from the analysis of the 
SMA data, except the total belt flux ($F_\text{belt}$), the total flux for 
the central peak ($F_\text{cen}$), and two position offsets 
$\{\Delta\alpha,\Delta\delta\}$. In essence, we fix the {\em shape} of the 
emission structure to that found from analysis of the SMA data, and we 
determine the 7~mm fluxes for the central component and for the belt component, 
allowing for the possibility of a small positional shift between the 
SMA and ATCA datasets.  We do not constrain $F_\text{belt}$ to be positive as for the SMA observations

\begin{figure}[ht]
\centerline{\psfig{file=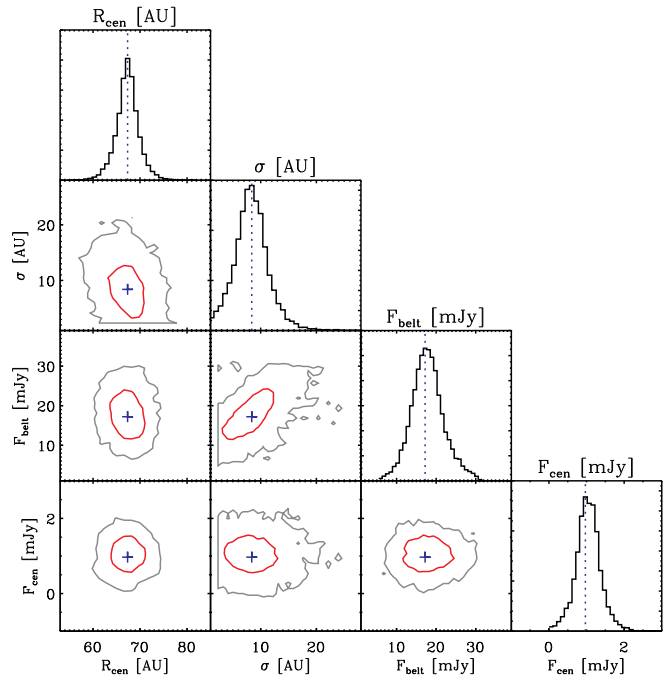,width=10cm,angle=0}}
\caption[]{\small 
A sample of the output from a run of $\sim10^4$ MCMC trials 
for the 4 best-fit Gaussian belt geometry parameters ($R_\text{cen}$, $\sigma$, 
$F_\text{belt}$, and $F_\text{cen}$). The diagonal panels show the 1D histogram for 
each parameter marginalized over all other parameters considered in the model.  For each parameter, 
the peak of each histogram is taken to be the best-fit value.
The remaining panels show contour plots of the $1\sigma$ (red) and $2\sigma$ 
(gray) regions for each pair of parameters, with the blue crosses marking the 
best-fit values. 
}
\label{fig:mcmc}
\end{figure}  

\subsection{Results of Model Fits}

\begin{figure}[ht]
\centerline{\psfig{file=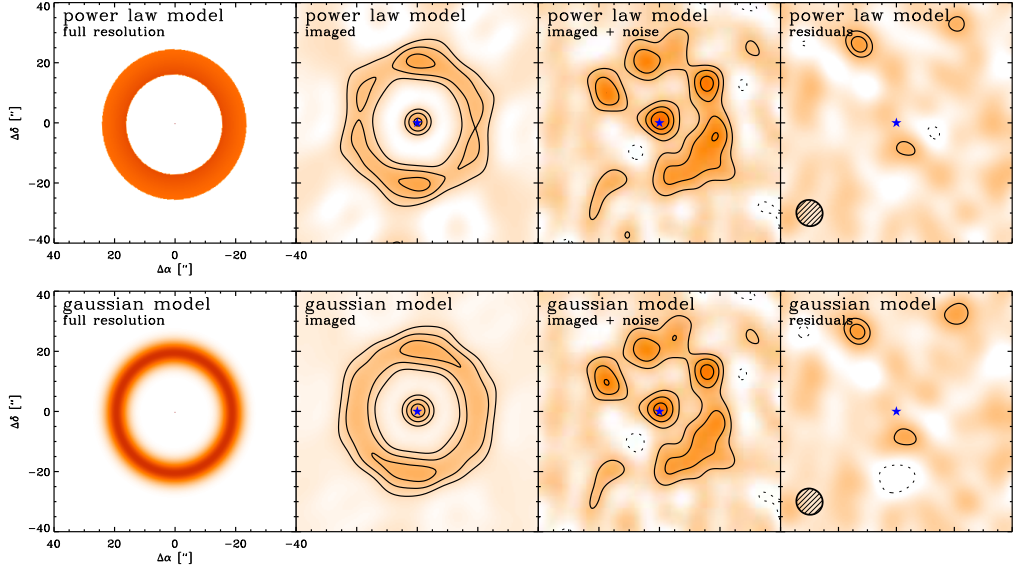,width=14cm,angle=0}}
\caption[]{\small 
Images of the best-fit models to the SMA 1.3~mm emission. 
\emph{(Upper): (left)} the best-fit power law disk model at full resolution, 
pixel scale $\sim 0\farcs2$ (0.8) AU, and
\emph{(center, left)} image of the best-fit power law disk model with no
noise, 
\emph{(center, right)} the best-fit power law disk model with simulated noise, 
and 
\emph{(right)} the imaged residuals from the power law disk model,
made using the same imaging parameters as in Figure~\ref{fig:SMA_ATCA}.
\emph{(Lower): (left)} the best-fit Gaussian disk model at full resolution, and
\emph{(center, left)} image of the best-fit Gaussian disk model,
\emph{(center, right)} image with simulated noise of the best-fit Gaussian disk model, and 
\emph{(right)} the imaged residuals from the Gaussian disk model, again
made using the same imaging parameters as in Figure~\ref{fig:SMA_ATCA}.
The contour levels are at 0.4 mJy beam$^{-1}$ ($2\sigma$) intervals 
in all panels.
The ellipse in the lower left corner of the residual images indicates the 
$9\farcs2 \times 7\farcs0$ (FWHM) synthesized beam size.
}
\label{fig:model_residual}
\end{figure}

Table~\ref{tab:sma_fits} lists the resulting best-fit 
parameter values and their $68\%$ uncertainties determined from the 
marginalized posterior probability distributions for both the power law and 
Gaussian belt models fit to the SMA 1.3~mm data.
Figure~\ref{fig:mcmc} shows a sample of the output for the main Gaussian
belt parameters, including the marginalized posterior probability distributions.
The $1\sigma$ and $2\sigma$ regions are determined by assuming normally 
distributed errors, where the probability that a measurement has a distance less 
than $a$ from the mean value is given by $\text{erf}\left(\frac{a}{\sigma\sqrt{2}}\right)$.
Both of these functional forms provide good fits 
to the observed visibilities, with reduced $\chi^2$ values of about $1.4$ 
(59,924 independent data points, 11 and 10 free parameters for the power law 
and Gaussian models, respectively) for each.  Figure~\ref{fig:model_residual}
shows the best-fit models of the 1.3~mm data in the image plane, at the
full resolution of the models, and imaged like the SMA data, both without
noise and with the noise level obtained by the observations, which results 
in patchy outer belt emission very similar to the SMA image in 
Figure~\ref{fig:SMA_ATCA}.
The imaged residuals are also shown in Figure~\ref{fig:model_residual},
and these are mostly noise (see Section~\ref{sec:azimuthal} for further 
discussion of the residuals from these axisymmetric models).

\begin{figure}[ht]
\centerline{\psfig{file=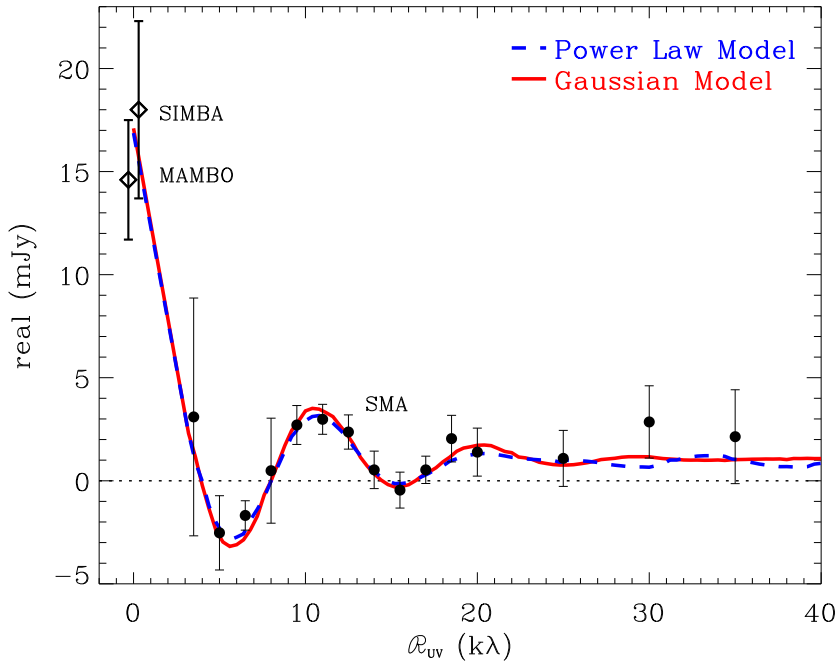,width=12cm,angle=0}}
\caption[]{\small The real part of the SMA 1.3~mm visibilities averaged 
in bins of deprojected (u,v) distance from the disk center, compared to 
the best-fit power law (dashed blue) and Gaussian (solid red) disk models
plus a central point-source.
The single dish MAMBO-2/IRAM \citep{les15} and SIMBA fluxes \citep{sch04} 
extrapolated from 1.2 mm to 1.3 mm are plotted at 
$\mathcal{R}_{uv} = 0$ k$\lambda$ (slightly offset from zero, for viewing clarity).
}
\label{fig:visfunction}
\end{figure}

A useful way of visualizing the interferometric data and model fits is 
through the deprojected visibility function, which takes advantage of 
(near) axisymmetry to reduce the dimensionality \citep{lay97}. In particular, 
the real part of the complex visibilities are averaged in concentric annuli 
of deprojected $(u,v)$ distance, $\mathcal{R}_{uv}$, from the center of the emission 
structure.  
Figure~\ref{fig:visfunction} shows this view of the SMA 1.3~mm data 
together with the best-fit power law and Gaussian belt models. The result 
is a function with a zero-crossing null and several subsequent oscillations. 
Although these
SMA observations are missing the short $(u,v)$ spacings needed to sample the 
peak of the visibility function, the overall shape matches nicely that 
expected for a narrow annulus of emission, plus a small and constant 
positive offset contribution from an unresolved central source.
Figure~\ref{fig:visfunction} also shows the single dish flux measurements 
at the zero-spacing of the deprojected visibility function (with small offsets 
from zero for clarity). It is remarkable that the simple axisymmetric belt
models of the SMA data appear to provide a very good estimate of the total 
flux, despite the lack of shorter baseline data.  This consistency 
provides indirect support for the basic model of the emission distribution.

Table~\ref{tab:atca_fits} lists the best-fit parameter values from the
modeling and their $68\%$ uncertainties for the 7~mm data.  This model 
again provides a good fit to the data, with reduced $\chi^2 = 1.32$ 
(151,976 independent data points, 4 free parameters). 
We note that the best-fit belt flux at 7~mm is positive, albeit with
low statistical significance; this positive value suggests the presence 
of emission below the detection threshold in individual beams in the image.
The position offsets between the ATCA and SMA data are small, consistent 
with zero.

\section{Discussion}
\label{sec:discussion}

We have performed interferometric observations of the $\epsilon$ Eridani 
system at 1.3~mm and 7~mm with the SMA and the ATCA, respectively, with baselines 
that sample to $\sim 4''$ (13~AU) resolution.  
The 1.3~mm image reveals emission from a resolved outer dust belt 
located $\sim 18\arcsec$ (60~AU) from the star, and a compact source 
coincident with the stellar position.  The 7~mm image shows only a 
central peak, detected with greater significance than in the 1.3~mm image.
We modeled the visibility data assuming two emission components, an outer belt 
with a power law or Gaussian radial surface brightness profile, and 
a central point source. Both functional forms provide good fits to the 
1.3~mm data, and we used the best-fit emission shape parameters to obtain 
constraints on the component flux densities from the 7~mm data. 
We now use the new information about these emission components to discuss 
implications for the debris disk and to compare to claims derived from 
previous millimeter and submillimeter observations with lower angular 
resolution.

\subsection{Outer Belt}
Several basic  properties of the outer emission belt are strongly 
constrained by the SMA 1.3~mm observations, including its flux, 
radial location and width, viewing geometry, and departures from axisymmetry. 
These new constraints bear on the possible presence of unseen planets in the 
system.

\subsubsection{Belt Flux}

The total flux density of the best-fit power law and Gaussian models, is 
constrained to be $F_\text{belt} = 16.9^{+3.9}_{-5.6}$ mJy and 
$17.2^{+5.1}_{-4.5}$ mJy, respectively.  These values are consistent with 
each other, within the uncertainties. They are also consistent with previous 
single-dish mapping measurements in this atmospheric window, accounting 
for minor differences in effective wavelength due to the broadband nature 
of bolometer detectors. \cite{les15} obtained a total flux density of 
$17.3 \pm 3.5$ mJy using MAMBO-2 on the IRAM 30-m telescope and \cite{sch04} 
measured a total flux density of $21.4 \pm 5.1$ mJy using SIMBA with the 
SEST 15-m telescope.  If we extrapolate these measurements using the 
spectral index of $\sim 2.14$ derived for $\epsilon$ Eridani at submillimeter 
wavelengths \citep{gas12data}, we find close agreement with the values 
obtained from the SMA analysis. In particular, extrapolating the IRAM/MAMBO-2 
observation to 1.3 mm gives $14.6 \pm 2.9$ mJy, while the older SIMBA 
observation gives $18.0 \pm 4.3$ mJy (see Figure~\ref{fig:visfunction}).

\subsubsection{Belt Location and Width}

For the best-fit power law model, the outer radius of the millimeter emission 
belt is determined to be $R_\text{out} = 80.2^{+3.0}_{-7.1}$ AU. Previous 
imaging studies of $\epsilon$ Eridani provide estimates of the outer 
radius between $70 - 90$ AU, in good agreement with this determination  
\citep{gre98,gre05,gre14,bac09}.  Additionally, we constrain the inner radius 
of the power law model, $R_\text{in} = 53.4^{+6.1}_{-5.0}$ AU.  This value 
agrees most closely with analysis of 160~$\mu$m {\em Herschel} observations 
by \citet{gre14}, which also suggest an inner radius of the outer belt 
of $\sim 54$ AU.  Other single-dish observations indicated that the belt 
extends further inward towards the star, $R_\text{in} = 35 - 40$ AU 
\citep{gre98,gre05,bac09}. The model fits to the SMA 1.3~mm data do not 
support such a wide belt with an inner radius so close to the star. 

The best-fit Gaussian model is characterized by a radial location, 
$R_\text{cen} = 64.4^{+2.4}_{-3.0}$ AU, and a width, 
$\sigma = 8.55^{+2.54}_{-3.46}$ AU, or 
FWHM $ = \Delta R = 2\sqrt{2\text{ln}(2)}\times \sigma = 20.2^{+6.0}_{-8.2}$~AU.  
These parameters are most directly comparable to the belt parameters derived 
by \cite{les15} using IRAM/MAMBO-2 observations at 1.2 mm with a 
telescope FWHM beam size of
$10\farcs7$; they fit a Gaussian shape to the disk emission radial profile
and obtain a central radius $R_\text{cen} = 57 \pm 1.3$ AU,
FWHM $= 12\arcsec \pm 1\arcsec$, and infer $8 \le \Delta R \le 22$ AU.  
This central radius is slightly 
smaller than that derived from the SMA data, but plausibly consistent 
within the mutual uncertainties. Since the lower limit on the width is 
narrower than implied by the fit to the SMA data at the 68\% confidence 
interval, we examine this potential discrepancy more closely.

The effect of changing the belt location ($R_\text{cen}$) and width 
($\sigma$) is dramatic on the null locations in the deprojected visibility 
function. 
We expressed our Gaussian model using a surface brightness profile of the 
form $I_\nu(r) = A\times\text{exp}[-((r-R_\text{cen})/\sqrt{2}\sigma)^2]$.  
The Fourier transform of a radially symmetric function like this can be 
expressed by a Hankel transform:

\begin{equation}
F(\rho) = 2\pi A\int_{0}^{\infty}I_\nu(r)J_0(\rho r)rdr,
\end{equation} 

\noindent where, $\rho = 2\pi\sqrt{u^2+v^2} = 2\pi\mathcal{R}_{uv}$.  
For a Gaussian ring, there is an exact solution to this integral involving an
infinite series of hypergeometric functions that can be evaluated numerically 
to find the exact locations of the visibility nulls. Fortunately, there is 
also an approximate solution to the Hankel transform using a generalized shift 
operator \cite[described in][]{bad09} that yields the values of the null 
locations to within 1\% of the exact solution, 

\begin{equation}
F(\rho) = 2\pi A\sigma^2\times\text{exp}[-(\rho\sqrt{2}\sigma)^2/4]\times J_0(\rho R_\text{cen}),
\end{equation}

\noindent
From this simple expression, we can see immediately that for a fixed belt 
width, decreasing the belt radius (inward towards the star) moves the 
zero-crossing null locations towards larger $\mathcal{R}_{uv}$.  For a fixed belt 
location, decreasing the belt width moves the zero-crossing null locations 
towards slightly smaller $\mathcal{R}_{uv}$ (and increases the amplitude of subsequent 
oscillations). 
The best-fit Gaussian model to the SMA data, $R_\text{cen} = 64$ AU, 
$\sigma = 8.6$ AU yields nulls at $\mathcal{R}_{uv} \approx 4$, 8, and 14 k$\lambda$. 
By comparison, a Gaussian model with $R_\text{cen} = 57$ AU and 
$\Delta R = 8$~AU ($\sigma \sim 4$ AU), at the lower limit of width suggested by \citet{les15}, 
results in nulls at $\mathcal{R}_{uv} \approx 5$, 9, and 16 k$\lambda$, which are 
significantly offset from the data. The differences between these 
fit results may stem from the oversimplified assumption of a strict Gaussian 
shape for the emission, perhaps exacerbated by deconvolution of the synthesized
beam resulting from the shift-and-add procedure used to restore the MAMBO map 
from the chopped observations of the $\epsilon$ Eridani field.

The presence of planets can affect the widths of planetesimal belts, through 
dynamical interactions. Given the best-fit Gaussian belt parameters, we can
constrain the fractional belt width of the $\epsilon$ Eridani debris disk
to $\Delta R/R = 0.31^{+0.09}_{-0.13}$.  This fractional width lies within 
the range of $0.1 \le \Delta R/R \le 0.4$ obtained by \cite{les15}.
For comparison, the classical Kuiper Belt in our Solar System appears radially 
confined between 
40 and 48 AU ($\Delta R/R \sim 0.18$), filling the region between the 3:2 
and 2:1 resonances with Neptune, likely the result of its outward migration 
\citep{hah05}. A narrow ring of millimeter emission in the Fomalhaut debris 
disk (FWHM $\sim16$ AU and $\Delta R/R \sim 0.1$) has been attributed to 
confinement by shepherding planets orbiting inside and outside the ring 
\citep{bol12}.  The best-fit value of the fractional width of the $\epsilon$ 
Eridani belt is wider than the Fomalhaut belt and the classical Kuiper Belt,
but not as wide as the belt surrounding the Sun-like star HD~107146, which 
also shows hints of a more complex radial structure \citep{ric15}.

The SMA 1.3~mm data do not place any strong constraints on the sharpness of 
the belt edges. This is evidenced by the comparable fit quality for both 
the sharp-edged power law and smoother Gaussian surface density profiles.  
Sharp edges in the underlying planetesimal distribution might be expected 
from planetary interactions, as regions within chaotic zone boundaries 
are rapidly cleared. For example, the sharp inner edge of the Fomalhaut 
debris disk seen in scattered light has long suggested sculpting by a planet 
\citep{qui06,chi09}. In contrast, ``self-stirred'' debris disks are not 
expected to show sharp edges.  In these models, the formation of Pluto-sized 
bodies initiate collisions that propagate outward to radii of several tens 
of AU over Gyr timescales \citep{ken08}. This process tends 
to produce a radially extended planetesimal belt, with an outwardly increasing 
gradient. The specific models of \citet{ken10} predict an $r^{7/3}$ profile 
for the belt optical depth.  Given the limits of the resolution and sensitivity
of the SMA data, model fitting does not provide a strong constraint on the 
power law gradient of the belt emission, $x = 1.92^{+0.18}_{-2.94}$.  However, 
if we take this best-fit exponent at face value and assume that the emitting 
dust is in radiative equilibrium with stellar heating, which gives a 
temperature gradient close to $T \propto r^{-0.5}$, then this value implies a 
rising surface density profile, $\Sigma \propto r^{2.4}$ (with large 
uncertainty).  This is similar to the rising surface density profile, 
$\Sigma \propto r^{2.8}$ found from millimeter observations of the 
AU~Mic debris disk \citep{mac13}, as well as the rising surface density 
towards the outer edge of the HD~107146 debris disk \citep{ric15}. 
Since the surface density of protoplanetary accretion disks decreases with 
radius, this small but growing sample of debris disks with rising gradients 
may point to support for self-stirred models of collisional excitation.

\subsubsection{Belt Viewing Geometry}

In addition to the belt parameters, the data place constraints on the disk
viewing geometry through the inclination and position angle parameters. 
The position angles from both the power law and Gaussian belt models are 
consistent with $PA = 0\degr$, as previous observations have found 
\citep{les15}.  The inclinations determined from the SMA models are
$i=17.9^{+10.2}_{-15.3}$ (power law) and 
$i=17.3^{+14.2}_{-14.2}$ (Gaussian), lower than but consistent with claims 
of $i \approx 25\degr$ from analysis of most other far-infrared and 
submillimeter images \citep{gre98,gre05,les15}.  \citet{gre14} fit a 
flat ring model to {\em Herschel} 160 $\mu$m data and obtain a higher 
inclination value, $i = 30\degr \pm 5\degr$, still compatible with the
fits to the SMA 1.3~mm data within 
the uncertainties. However, this difference could be a sign of background 
confusion affecting the inferences from the far-infrared images, or perhaps 
a wavelength dependence of this parameter due to the emission sampling 
different grain size populations.
To assess whether or not such a higher inclination could affect the other 
belt parameters in the SMA analysis, we fixed $i=30\degr$ and re-ran the 
MCMC model fitting procedure; the best-fit parameters are hardly changed 
($< 3\%$). 

\subsubsection{Limits on Belt Stellocentric Offset}

The model fits show no significant centroid offset between the belt and 
central component,
as might result from the secular perturbations of a planet in an eccentric 
orbit interior to the belt. For planet induced eccentricities, 
the displacement of the belt centroid from the star should be $\sim ae$, where 
$a$ is the semi-major axis of the belt and $e$ is the forced eccentricity 
\cite[e.g.][]{chi09}.  Our modeling allows us to place a robust $3\sigma$ upper
limit on the displacement, $\Delta r_\text{cen} \lesssim 2\farcs7 = 8.7$ AU.  
Based on a possible far-infrared north-south flux asymmetry attributed to 
pericenter glow (enhanced emission at periapse), \cite{gre14} raise 
the possibility of an additional planet in the $\epsilon$ Eridani system 
with semi-major axis within the outer belt $a = 16-54$~AU and eccentricity 
$e \approx 0.03 - 0.3$.  Given the constraint on the centroid offset from the 
SMA data, the presence of a giant planet on a wide orbit of several 10's of AU 
with a large eccentricity, $e \gtrsim 0.1$ is disfavored. 
This is in accord with direct imaging constraints at infrared wavelengths 
that preclude planets of about 1 Jupiter mass beyond 30 AU \citep{jan15}.
However, the effects
of a Uranus or Neptune-like planet with lower orbital eccentricity is still 
easily accommodated within the limits.

\subsubsection{Limits on Belt Azimuthal Structure}
\label{sec:azimuthal}

The azimuthal structure of the $\epsilon$ Eridani debris disk has been the 
subject of much debate. The first JCMT/SCUBA 850 $\mu$m image of the disk 
showed a non-uniform brightness distribution with several peaks of modest 
signal-to-noise ratio \citep{gre98}. Follow-up JCMT/SCUBA observations from 
up to 5 years later suggested that three of the peaks in the original image 
appear to move with the stellar proper motion, and in fact showed tentative 
evidence of counterclockwise rotation of $\sim 1\degr$ yr$^{-1}$ \citep{gre05}.
Several other peaks did not appear to move with 
the star and were presumed to be background features.  
\citet{les15} claim from IRAM/MAMBO-2 observations that the disk
shows a similar azimuthal structure, with four peaks, in the northeast, 
southeast, southwest, and northwest. However, {\em CSO} observations at 
350~$\mu$m did not confirm the same peaks \citep{bac09}, and {\em Herschel} 
observations at 250~$\mu$m (albeit at lower resolution) show a relatively 
smooth emission distribution, with the southern end $\sim 10\%$ 
brighter than the northern end \citep{gre14}.

The interest in determining the robustness of the $\epsilon$ Eridani clump 
structure stems from the suggestion that the outward migration of a planet  
could trap planetesimals outside of its orbit in mean motion resonances.  
A variety of numerical simulations show that the pattern of clumps observed 
in a disk depends on the planet mass and the resonances involved 
\cite[e.g.][]{kuc03,wya03,del05}.  
Thus, the emission morphology of a debris belt can be diagnostic of the 
presence of a planet and its migration history.  However, other numerical 
simulations suggest that all azimuthal asymmetries should be effectively 
erased by collisions within debris disks as dense as $\epsilon$ Eridani 
\citep{kuc10}, which would imply that any clumps are spurious, or perhaps
background sources.  Background sources seem particularly problematic to
the east of $\epsilon$ Eridani in wide field {\em Herschel} submillimeter 
images, which likely contributed to confusion at earlier epochs when the 
disk was superimposed on them.

\begin{figure}[ht]
\centerline{\psfig{file=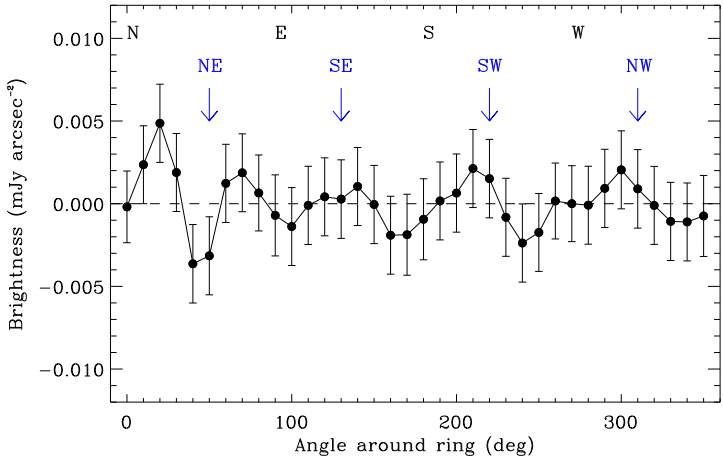,width=10cm,angle=0}}
\caption[]{\small Azimuthal profile of the residual emission after subtracting 
our best-fit Gaussian model from the 1.3 mm continuum emission.   
The four clumps discussed by \cite{gre05} and \cite{les15} are marked by the blue arrows.
}
\label{fig:azprofile}
\end{figure} 

The SMA 1.3 mm data probes structure at higher angular 
resolution than the previous single dish observations. Moreover, 
the interferometer naturally provides spatial filtering that serves to 
highlight the presence of any compact emission peaks.
After removing azimuthally symmetric models from the SMA data, any 
significant azimuthal structure should be readily apparent in the imaged
residuals. Figure~\ref{fig:azprofile} shows the azimuthal profile of the 
residual image obtained after subtracting the best-fit Gaussian belt model 
from the data.  Each point represents the mean brightness calculated in a 
small annular sector with opening angle of $10\degr$ and radial range of 
$10\arcsec$ to $30\arcsec$.  Uncertainties are the image rms noise divided by 
the square root of the number of beams in each sector. 
The only potentially significant feature is a peak at $20\degr$, which 
is visible at the $\sim 4\sigma$ level in the residual images of 
Figure~\ref{fig:model_residual} (right panels).  
The locations of the previously claimed 
four clumps \citep{gre05,les15} are marked with arrows
in Figure~\ref{fig:azprofile}.  While the azimuthal profile of the 
imaged residuals shows roughly four low significance peaks, these are not 
aligned well with the previously claimed clumps, and the separations
cannot be readily attributed to the previously claimed rotation. However,
the signal-to-noise of these residuals is still lacking. The signature of the four clumps discussed in \cite{les15}, if they do exist, have been weakened in the fitting procedure by using an azimuthally uniform ring as the model. A more definitive 
statement about low level clumps will require high resolution observations 
with higher sensitivity.

\begin{figure}[ht]
\centerline{\psfig{file=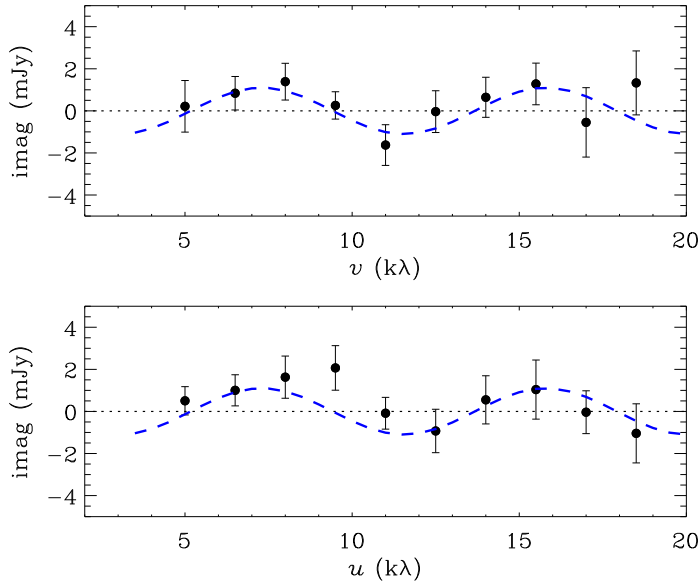,width=10cm,angle=0}}
\caption[]{\small Complex visibilities of the 1.3 mm emission binned along the $v$-axis \emph{(top)} and the $u$-axis \emph{(bottom)}.  Model visibilities generated by inserting a 1 mJy point source at the position of the $4\sigma$ peak seen in the residual map are shown by the blue dashed lines.
}
\label{fig:imag}
\end{figure}

The imaginary part of the visibilities is very sensitive to the presence 
of asymmetries in the emission structure. Indeed, the effect of the 
marginally significant peak in the residual images is apparent.
Figure~\ref{fig:imag} shows the imaginary part of the visibilities 
binned along the $u$-axis and along the $v$-axis of the Fourier plane.
For a symmetric structure, the imaginary part of the visibilities should be 
zero; however, both of these views show hints of a low amplitude, sinusoidal 
oscillation. This structure in the visibilities arises naturally from 
a single, offset emission peak. The Fourier transform of a point source 
with amplitude, $A$, offset from the center of the image in the $u$ and $v$ 
directions by $x_0$ and $y_0$, respectively, is given by 
$G(u,v) = A \times \text{exp}[-2\pi(x_0u+y_0v)]$
(making use of the shift theorem, whereby  a shift in the position of a 
function by an amount $x_0$ corresponds to a phase change in its Fourier 
transform by exp$(i2\pi x_0 u)$, e.g. \citet{ise13}); the imaginary part 
of this expression is 
$\mathcal{I}(G(u,v)) = A \times \text{sin}[2\pi(x_0u+y_0v)]$. 
That is, a point source offset from $(0,0)$ introduces a simple 
sinusoidal oscillation in the imaginary part of the visibilities.
As shown in Figure~\ref{fig:imag}, if we insert a 1~mJy point source 
at the location of the $4\sigma$ peak in the residual image
($x_0 \approx 15\arcsec$, $y_0 \approx 25\arcsec$) into the visibilities,
the resulting imaginary visibility curve matches very well the shape and 
scale of the residuals. 
Thus, this single residual feature can account for most of the 
structure in the imaginary visibilities, or most of the detected asymmetry.  
This feature does not match up with any of the clumps previously identified 
in single dish images. Also, it is positioned clockwise from the northeast 
feature, which is not consistent with the 
counterclockwise rotation suggested by \cite{gre05}.  The exact nature of 
this marginally significant peak is uncertain.
If it is a background galaxy, then the proper motion 
of the star will move the debris disk away from it, and this should become 
evident in interferometric observations at future epochs.  
An extragalactic source should also appear compact at the arcsecond level.  
Recent deep ALMA surveys \citep{hod13,kar13} have built up statistics 
describing the number counts of submillimeter galaxies expected in a given 
field of view.  At 1.3~mm, the expected number of sources with flux $\ge 1$~mJy
expected in the $52\arcsec$ primary beam of the SMA is $0.41^{+0.25}_{-0.15}$ 
\citep{ono14}. Hence the presence of a background submillimeter galaxy in 
the image at this flux level is not a rare event. 
Verification and characterization of any asymmetric structure in the 
millimeter emission from the belt requires observations with higher resolution 
and sensitivity.

\subsubsection{Belt Spectral Index}
The long wavelength spectral index of the continuum belt emission reflects the 
underlying dust opacities and provides a constraint on the size distribution 
of the emitting grains in the debris disk \citep{ric12}, which can be related 
to collisional models.  

For optically thin dust emission, the flux density is given by 
$F_\nu \propto B_\nu(T_\text{dust})\kappa_\nu M_\text{dust}/d^2$, 
where $B_\nu(T_\text{dust})$ is the Planck function at the dust temperature 
$T_\text{dust}$, $\kappa_\nu \propto \nu^\beta$ is the dust opacity, expressed 
as a power law at these long wavelengths, $M_\text{dust}$ is the dust mass, 
and $d$ is the distance.  At long millimeter wavelengths, for sufficiently
high temperatures, the Planck function reduces to the Rayleigh-Jeans 
approximation $B_\nu(T_\text{dust}) \propto \nu^{\alpha_\text{Pl}}$ 
with $\alpha_\text{Pl} = 2$. 
Thus, $\alpha_\text{mm} \approx \alpha_\text{Pl}+\beta$, 
where $\alpha_\text{mm}$ is the millimeter spectral index.  
\cite{dra06} derived a relation between $\beta$, the dust opacity power law 
index, and $q$, the grain size distribution parameter: 
$\beta \approx (q-3)\beta_s$, where $\beta_s = 1.8 \pm 0.2$ is the dust opacity
spectral index in the small particle limit for interstellar grain materials, 
valid for $3 < q < 4$ and size distributions that follow a power law over a 
broad enough interval. Combining these relationships provides a simple 
expression for the slope of the grain size distribution, $q$, as a function 
of $\alpha_\text{mm}$, $\alpha_\text{Pl}$, and $\beta_s$:
$q = (\alpha_\text{mm} - \alpha_\text{Pl})/\beta_s + 3$.

For $\epsilon$~Eridani, 
the fit to the ATCA 7~mm data places an upper limit on 
the belt flux density of $F_\text{belt} < 310$ $\mu$Jy ($3\sigma$).
Combining this 7~mm limit with the SMA 1.3~mm measurement provides a 
long lever arm in wavelength that largely overcomes systematic uncertainties 
and constrains the millimeter spectral index, $\alpha_\text{mm} > 2.39$. 
This limit on the spectral index results in a limit on the grain size 
distribution power-law index $q > 3.22$.  

The derived limit on the grain size distribution power-law index is consistent 
with the classical prediction of $q = 3.5$ for a steady-state collisional 
cascade \citep{doh69}.  \citet{ric12} obtained a similar result from analysis 
of the millimeter spectrum of the Fomalhaut debris disk, $q = 3.48 \pm 0.14$.  
This standard collisional cascade assumes that collisions in the disk occur 
between bodies of identical tensile strength and velocity dispersion, 
regardless of size.  \cite{pan12} revisited the theory by
relaxing the assumption of a single velocity dispersion and solving 
self-consistently for a size-dependent velocity distribution in steady-state.  
This more complex analysis yields steeper size distributions, with 
$q \approx 4$. Since the $\epsilon$~Eridani spectral index is only a 
lower limit, a steeper grain size distribution cannot be ruled out. 
We note that the best-fit value for the 7~mm flux density 
($F_\text{belt} = 110~\mu$Jy) yields $q = 3.55 \pm 0.36$, still shallower than 
predicted by models with size-dependent velocity dispersions.  
This result, and the more robust measurement for Fomalhaut, do not support 
the steeper size distributions predicted by the collisional models with 
velocity distribution variations.  But spectral indices need to be determined 
for a much larger sample of debris disks to draw a definitive conclusion.

\subsection{Central Component}

A central source is clearly detected in the SMA 1.3~mm image and ATCA 7~mm 
image, coincident with the position of the star at the time of observation. 
The source size is below the resolution limit of these observations, most 
clearly evidenced by the lack of fall off at long baselines 
in Figure~\ref{fig:visfunction}.  The 1.3~mm flux density of this source is 
$F_\text{cen} = 1.08^{+0.19}_{-0.41}$~mJy.
\cite{les15} report a similar value in MAMBO-2/IRAM 1.2~mm data, detecting 
a central source with flux density $1.2 \pm 0.3$ mJy. These measurements 
are only marginally compatible with expectations for the stellar photosphere 
at these long wavelengths.  The effective temperature of $\epsilon$ Eridani 
is $T_\text{eff} = 5039 \pm 126$~K \citep{bai12}, and a Kurucz stellar 
atmosphere model \cite[see][]{bac09} predicts a 1.3~mm flux density of 
0.53~mJy (with 2\% uncertainty).  The ATCA 7~mm flux density of this same 
central source is $F_\text{cen} = 66.1^{+6.9}_{-10.5}$ $\mu$Jy, substantially 
in excess of the stellar photoshere model flux prediction of 18~$\mu$Jy.  
The central source persists at a consistent intensity in all of the 8 days
of ATCA observations, showing no significant variability.  The mean flux density
is $66.5 \pm 9.5$ $\mu$Jy and $65.9 \pm 7.0$ $\mu$Jy for the June and August 
observations, respectively.

In principle, the central 1.3~mm excess emission could be explained by 
thermal dust emission from a warm inner belt. This small 1.3~mm excess, 
together with {\em Spitzer} 24~$\mu$m and {\em Herschel} 70 and 160~$\mu$m 
inner excesses, are consistent with emission from a $70-100$~K blackbody, 
similar to previous inferences by \citet{bac09} and \citet{gre14} from 
the infrared spectrum alone.  For reasonable grain sizes, this blackbody 
emission corresponds to a (very narrow) inner dust belt at $2-10$~AU, 
consistent with the size constraint from the SMA observations. However, this 
same blackbody model produces negligible emission at 7~mm.  While previous
infrared measurements indicate that there is clearly some warm dust present in the system
(unresolved in our observations), no inner dust belt scenario 
can also match the substantial 7~mm excess from the ATCA observations.

We consider it likely that the unresolved excess emission from the central 
source arises from an additional {\em stellar} component, either an ionized 
corona or chromosphere. The absence of variability on the month to month 
timescale suggests a thermal origin.
In particular, the millimeter wavelength emission from $\epsilon$~Eridani 
is reminiscent of the nearby solar-type stars $\alpha$ Cen A and B (spectral 
types G2V and K2V) recently reported by \citet{lis15} and attributed to 
heated plasma, similar to the Sun's chromosphere.  Following \citet{lis13}, 
we calculate the Planck brightness temperature for $\epsilon$ Eridani at 
1.3~mm and 7~mm, assuming the photospheric radius of the star is sufficiently 
similar at optical and radio wavelengths to introduce negligible errors.  
Optical interferometry of $\epsilon$ Eridani gives a precise measure of the 
stellar radius, $R_\text{phot} = 0.74 \pm 0.01 R_\odot$ \citep{bai12}. At 1.3~mm, this 
radius and the excess emission implies $T_\text{B} = 7800 \pm 1400$ K, 
somewhat higher than the optical effective temperature.  At 7~mm, however, 
$T_\text{B} = 13000 \pm 1600$ K, very much in excess of the photospheric 
prediction. Indeed, the $\epsilon$~Eridani emission follows the same trend 
with increasing wavelength as found for $\alpha$ Cen A and B from ALMA 
observations.  \citet{lis15} measure spectral indices between 0.87~mm and 
3.1~mm of $1.62$ and $1.61$ for $\alpha$ Cen A and B, respectively. 
From the SMA and ATCA data, the spectral index of the central component of 
$\epsilon$~Eridani between 1.3~mm and 7~mm is very similar, $1.65 \pm 0.23$ 
(where we have added in quadrature the $\sim10\%$ flux scale uncertainties 
at both wavelengths with the $1\sigma$ errors from model fits). The stellar 
spectrum clearly starts to deviate strongly from a simple optically thick 
photosphere with a Rayleigh-Jeans spectral index of 2.0, and the contrast 
between the photosphere and the putative chromosphere increases at longer 
wavelengths. While observations of $\epsilon$~Eridani at centimeter 
wavelengths have so far provided only upper limits, $< 80~\mu$Jy at 3.6~cm 
\citep{gud92} and $<105~\mu$Jy at 6~cm \citep{bow09}, much more sensitive 
observations are now possible with the upgraded Karl G. Jansky Very Large 
Array, and this would be useful to help constrain the plasma properties.

\section{Conclusions}

We present SMA 1.3~mm and ATCA 7~mm observations of $\epsilon$ Eridani, 
the first millimeter interferometric observations of this nearby debris disk 
system, probing to $4''$ (13~AU) scales. These observations resolve the 
outer dust emission belt surrounding the star, and they reveal a compact 
emission source coincident with the stellar position.  We use MCMC techniques 
to fit models of the emission structure directly to the visibility data in 
order to constrain the properties of the two components. 
The main results are:

\begin{enumerate}

\item The outer belt is located precisely and resolved radially. 
Gaussian and power law emission profiles each fit the SMA 1.3~mm data 
comparably well. For the best-fit Gaussian model, the belt radial location 
is $R_\text{cen} = 64.4^{+2.4}_{-3.0}$~AU and 
$\text{FWHM} = 20.2^{+6.0}_{-8.2}$~AU, corresponding to a fractional belt 
width $\Delta R/R = 0.31^{+0.09}_{-0.13}$. This width is at the high end 
of inferences from previous single dish millimeter observations,
and wider than the classical Kuiper Belt in our Solar System. 
    
\item The outer belt shows no evidence for significant azimuthal 
structure that might be attributed to gravitational sculpting by planets. 
After subtracting a symmetric model from the SMA 1.3~mm data, imaging 
shows only one low significance peak, and its location does not correspond 
to any clumps identified in previous millimeter and submillimeter observations 
of $\epsilon$~Eridani.  The presence of this feature is consistent with 
source counts for the extraglactic background in the field of view. 
In addition, the SMA 1.3~mm data constrains any centroid offset of the 
belt from the star to $<9$~AU, which limits the presence of giant planet 
perturbers on wide and eccentric orbits in the system.

\item A central source coincident with the star is clearly detected in 
both the SMA 1.3~mm image and the ATCA 7~mm image, and the flux densities 
of this source exceed extrapolations from shorter wavelengths for the 
stellar photosphere. While the excess is marginal at 1.3~mm, it is highly
significant-- about a factor of three-- at 7~mm.  
The stellar spectrum clearly departs from an optically thick photosphere 
at these long wavelengths, with spectral index $1.65\pm0.23$ between 1.3~mm
and 7~mm.  This spectrum cannot be explained by an inner warm dust belt 
and plausibly results from heated plasma in a stellar chromosphere, by 
analogy with the Sun and $\alpha$~Cen. The 
high brightness temperature at 7~mm of $13000\pm1600$~K for a source of 
stellar size lends additional credence to this conclusion.
   
\item Combining the SMA 1.3~mm measurement of the belt flux density with 
the ATCA 7~mm upper limit constrains the spectral index of the emission, 
$\alpha_\text{mm} > 2.39$. For conventional assumptions about the dust
grains,  this spectral index corresponds to a limit on 
the slope of the power law grain size distribution in the belt, $q > 3.22$, 
consistent with the classical prediction of $q = 3.5$ for a self-similar
steady-state collisional cascade. This slope is also consistent with the 
steeper distributions predicted by collisional models that allow for 
size-dependent velocities and strengths. 
  
\end{enumerate}

These SMA and ATCA millimeter wavelength observations provide the highest 
resolution view of the outer dust belt surrounding $\epsilon$~Eri at the 
longest wavelengths to date. But deeper observations are still needed to 
measure radial gradients in the debris disk and to reveal substructure due 
to planets, if present, in order to further constrain scenarios for the 
evolution of planetesimals surrounding this very nearby star.

\acknowledgements
M.A.M acknowledges support from a National Science Foundation 
Graduate Research Fellowship (DGE1144152) and from the Swinburne Centre for Astrophysics \& Supercomputing.
D.J.W. thanks the Swinburne Visiting Researcher Scheme.
S.T.M. acknowledges the support of the visiting professorship scheme from the Universit{\'e} Claude Bernard Lyon 1.
The Submillimeter Array is a joint project between the Smithsonian 
Astrophysical Observatory and the Academia Sinica Institute of Astronomy 
and Astrophysics and is funded by the Smithsonian Institution and the 
Academia Sinica.
We thank Mark Gurwell and Scott Paine for discussions about the 
Submillimeter Array primary beam shape.

\bibliography{References}

\begin{appendices}
\section{Primary Beam Structure of the SMA}
\end{appendices}

The SMA is composed of eight essentially identical antennas, each 6 meters 
in diameter. The SMA primary beam is thus the power pattern of one antenna. 
While the primary beam shape is often assumed to be a simple Gaussian, the
actual shape is determined by illumination with a 10 dB taper at the edge 
of the primary dish, as well as blockage due to the secondary mirror. With
these considerations, the beam power as a function of offset (in arcseconds) 
from the beam center is given by 

\begin{equation}
P = \left[\int^{R_s}_{R_p} 2\pi rJ_0\left(\frac{2\pi rx}{\lambda}\right)J_0\left(\frac{1.840839r}{R_p}\right)\right]^2
\label{eqn:primarybeam} 
\end{equation}

\noindent where, $R_p$ is the radius of the primary dish in meters, $R_s$ is 
the radius of the secondary dish in meters, and $x$ is the offset from the 
dish center in radians.  A complete profile of the beam power can be 
built up using Equation~\ref{eqn:primarybeam} at discrete offset positions. 
Note that this expression does not take into account additional practical 
factors, such as receiver alignment, pointing jitter, and departures from
perfect focus, which act to distort the primary beam shape.

Since the emission extent of the $\epsilon$~Eridani disk is comparable to 
the half power size of the SMA primary beam pattern, we constructed a 
complete beam model for use in our modeling procedure.  The FWHM of accurate 
beam models for the LSB (218.9~GHz) and USB (230.9 GHz) are $53\farcs6$ 
and $50\farcs8$, respectively. For comparison, the FWHM for a uniformly 
illuminated circular aperture antenna is given by $1.22 \lambda/D_A$, 
where $D_A$ is the antenna diameter.  For the SMA antennas, this predicts 
$50\farcs4$ and $47\farcs4$, for the LSB and USB, respectively, narrower 
than the FWHM of the accurate beam models.  Note that tasks within the miriad 
software package assume a Gaussian beam for the SMA with a FWHM given by a 
uniformly illuminated circular aperture.

\begin{deluxetable}{c c c c c}
\tablecolumns{7}
\tabcolsep0.1in\footnotesize
\tabletypesize{\small}
\tablewidth{0pt}
\tablecaption{Submillimeter Array Observations of $\epsilon$ Eridani}
\tablehead{
\colhead{Observation} & 
\colhead{Array} & 
\colhead{Projected} & 
\colhead{HA} &
\colhead{225 GHz atm.} \\
\colhead{Date} & 
\colhead{Config.} & 
\colhead{Baselines (m)} & 
\colhead{Range} &
\colhead{Opacity$^\text{a}$}
}
\startdata
2014 July 28 & Subcompact & $6 - 35$ & $-3.6, 3.3$ & $0.09$\\
2014 July 29 & Subcompact & $6 - 35$ & $-3.7, 2.8$ & $0.07$\\
2014 July 30 & Subcompact & $6 - 35$ & $-3.5, 7.1$ & $0.08$\\
2014 Aug 5 & Subcompact & $6 - 35$ & $-3.5, 2.2$ & $0.12$\\
2014 Nov 19 & Subcompact & $6 - 56$ & $-3.6, 4.3$ & $0.07$
\enddata
\tablecomments{$^\text{a}$ characteristic value for the track measured at 
the nearby Caltech Submillimeter Observatory. The LO frequency for all observations was 225.5 GHz.}
\label{tab:sma}
\end{deluxetable}

\begin{deluxetable}{c c c c c}
\tablecolumns{7}
\tabcolsep0.1in\footnotesize
\tabletypesize{\small}
\tablewidth{0pt}
\tablecaption{Australia Telescope Compact Array Observations of $\epsilon$ Eridani}
\tablehead{
\colhead{Observation} & 
\colhead{Array} & 
\colhead{Projected} & 
\colhead{HA} &
\colhead{Seeing rms} \\
\colhead{Date} & 
\colhead{Config.} & 
\colhead{Baselines (m)} & 
\colhead{Range} &
\colhead{(microns)$^\text{a}$}
}
\startdata
2014 June 25 & H168 & $36 - 180$ & $-3.6, 3.6$ & $150$\\
2014 June 26 & H168 & $36 - 180$ & $-3.6, 3.6$ & $60$\\
2014 June 27 & H168 & $31 - 180$ & $-4.0, 3.6$ & $70$\\
2014 June 28 & H168 & $31 - 180$ & $-4.1, 0.0$ & $250$\\
2014 Aug 2 & H75 & $22 - 84$ & $-4.7, 3.9$ & $80$\\
2014 Aug 3 & H75 & $22 - 84$ & $-4.4, 4.0$ & $80$\\
2014 Aug 4 & H75 & $22 - 84$ & $-4.6, 4.0$ &  $150$\\
2014 Aug 5 & H75 & $22 - 84$ & $-3.8, 4.0$ & $130$
\enddata
\tablecomments{$^\text{a}$ characteristic value for the track measured by
the ATCA seeing monitor, an interferometer on a 230 m east-west baseline that tracks the 30.48 GHz beacon on the geostationary communications satellite, OPTUS-B3, at an elevation of $60\degr$. \citep{mid06}. The LO frequency for all observations was 44 GHz.}
\label{tab:atca}
\end{deluxetable}

\begin{deluxetable}{l l c c }
\tablecolumns{4}
\tabcolsep0.1in\footnotesize
\tabletypesize{\small}
\tablewidth{0pt}
\tablecaption{SMA Model Parameters}
\tablehead{
\colhead{Parameter} & 
\colhead{Description} & 
\colhead{Power Law} & 
\colhead{Gaussian}\\ 
\colhead{} & 
\colhead{} & 
\colhead{Best-Fit} & 
\colhead{Best-Fit} 
}
\startdata
$F_\text{belt}$ & Belt flux density (mJy) & $16.9 (+3.9, -5.6)$ & $17.2 (+5.1, -4.5)$ \\
$F_\text{cen}$ & Central source flux (mJy) & $1.08 (+0.19,-0.41)$ & $1.06 (+0.34, -0.34)$ \\
$R_\text{in}$ & Belt inner radius (AU) & $53.4 (+6.1, -5.0)$ & $-$ \\
$R_\text{out}$ & Belt outer radius (AU) & $80.2 (+3.0, -7.1)$ & $-$ \\
$x$ & Belt radial power law index & $1.92 (+0.18, -2.94)$ & $-$\\
$R_\text{cen}$ & Belt center radius (AU) & $-$ & $64.4 (+2.4, -3.0)$ \\
$\sigma$ & Belt width (AU) & $-$ & $8.55 (+2.54, -3.46)$ \\
\hline
$i$ & Belt inclination $(^\circ)$ & $17.9 (+10.2, -15.3)$ &  $17.3 (+14.2, -14.2)$ \\
$PA$ & Belt position angle $(^\circ)$ & $3.42 (+23.3, -23.4)$ & $ 1.66 (+6.70, -6.70)$  \\
\hline
$\Delta\alpha$ & R.A. offset of belt center $(\arcsec)$ & $0.01 (+0.65, -1.03)$ & $0.02 (+0.80, -0.80)$ \\
$\Delta\delta$ & Decl. offset of belt center $(\arcsec)$ & $1.63 (+0.86, -0.86)$ & $1.63 (+0.70,-1.01)$ \\
$\Delta\alpha_\text{star}$ & R.A. offset of star from belt center $(\arcsec)$ & $-1.18 (+0.65, -1.40)$ & $-1.22 (+0.87, -1.26)$ \\
$\Delta\delta_\text{star}$ & Decl. offset of star from belt center $(\arcsec)$ & $0.11 (+0.90, -1.29)$ & $0.11 (+1.10, -1.10)$ \\
\enddata
\label{tab:sma_fits}
\end{deluxetable}

\begin{deluxetable}{l l c}
\tablecolumns{3}
\tabcolsep0.1in\footnotesize
\tabletypesize{\small}
\tablewidth{0pt}
\tablecaption{ATCA Model Parameters}
\tablehead{
\colhead{Parameter} & 
\colhead{Description} & 
\colhead{Best-Fit} 
}
\startdata
$F_\text{belt}$ & Belt flux density ($\mu$Jy) & $110. (+65., -117.)$ \\
$F_\text{cen}$ & Central source flux ($\mu$Jy) & $66.1 (+6.9,-10.5)$ \\
\hline
$\Delta\alpha$ & R.A. offset of belt center $(\arcsec)$ & $-0.87 (+0.72, -0.87)$  \\
$\Delta\delta$ & Decl. offset of belt center $(\arcsec)$ & $0.38 (+0.86, -0.62)$  \\
\enddata
\label{tab:atca_fits}
\end{deluxetable}

\end{document}